\documentstyle[prl,twocolumn,aps,tighten,floats,epsfig]{revtex}
\begin{document}
\twocolumn[\hsize\textwidth\columnwidth\hsize\csname @twocolumnfalse\endcsname

\title{Novel Coexisting Density Wave Ground State in Strongly Correlated,
 Two-Dimensional Electronic Materials}
\maketitle
\vskip 0.25 truein
\centerline{S. Mazumdar$^*$, R. T. Clay$^\dagger$ and D. K. Campbell$^\dagger$}
\vskip 1pc
\centerline{$^*$Department of Physics, University of Arizona, Tucson, AZ 85721}
\centerline{$^\dagger$Department of Physics, University of Illinois, Urbana, IL
61801}
\vskip 0.25 truein]
\noindent {\bf Two-dimensional (2D) strongly correlated electron systems
  underlie many of the most important phenomena in contemporary
  condensed matter physics, including the Quantum Hall Effect (QHE)
  \cite{qhe}, ``high $T_c$'' superconductivity
  \cite{hightc}, and possible exotic conducting
  states in silicon MOSFETs \cite{sarachik}.
  We demonstrate the existence of yet another exotic
  ground state in strongly correlated, 2D electronic
  materials: a novel, insulating bond-order/charge density wave state (BCDW)
  in the commensurate 1/4-filled band
  that persists for all anisotropies within the 2D lattice,
  in contradiction to the non-interacting electron prediction
  of the vanishing of density waves in 2D for non-1/2-filled bands.
The persistence of the BCDW in the 2D lattice is a 
consequence of strong electron-electron (e-e) interaction and the resultant 
``confinement,'' a concept recently widely debated \cite{Anderson}. 
Our results have implications
  for experiments in the organic charge transfer solids (CTS),
  where they explain the observation of a ``mysterious''
  coexistence of density waves \cite{Pouget,Toyota}, clarify
  the optical conductivity of the ``metallic'' state \cite{schwartz},
  and suggest an approach to the observed organic
  superconductivity.}
\vskip 1pc
We consider a 2D lattice, consisting of
coupled chains of strongly correlated electrons, 
described by the quasi-2D extended Peierls-Hubbard Hamiltonian
$$H = H_0 +H_{ee} + H_{inter} \eqno(1a)$$
$$H_0 = -\sum_{j,m,\sigma}[t-\alpha(\Delta_{j,m})]B_{j,j+1,m,m,\sigma}
+ \beta\sum_{j,m}v_{j,m}n_{j,m} $$
$$+ K_1/2\sum_{j,m}(\Delta_{j,m})^2 + K_2/2 \sum_{j,m}v_{j,m}^2 \eqno(1b)$$
$$H_{ee}=U\sum_{j,m}n_{j,m,\uparrow}n_{j,m,\downarrow} +
 V\sum_{j,m}n_{j,m}n_{j+1,m} \eqno(1c)$$
$$H_{inter} = 
-t_{\perp}\sum_{j,m,\sigma}
B_{j,j,m,m+1,\sigma} \eqno(1d)$$

In the above, $j$ is a site index, $m$ is a chain index, $\sigma$ is
spin, and we assume a rectangular lattice \cite{xysym}.
As $t_{\perp}$ varies from 0 to $t$, the electronic
properties vary from quasi-1D to quasi-2D, modeling
a wide range of materials. Each site
is occupied by a molecular unit, the displacement of
which from equilibrium is described by  $u_{j,m}$
(with $\Delta_{j,m}=(u_{j,m}-u_{j+1,m})$), $v_{j,m}$ is an intramolecular
vibration, and 
$B_{j,k,\ell,m,\sigma} \equiv
[c_{j,\ell,\sigma}^\dagger c_{k,m,\sigma} + h.c.]$, where 
$c_{j,\ell,\sigma}^\dagger$ is a Fermion operator.
We treat the phonons in the adiabatic
approximation and are interested in unconditional broken symmetry solutions
that occur for 
electron-phonon (e-ph)
couplings $(\alpha, \beta) \rightarrow 0^+$.
Although we limit our explicit analysis of e-e interactions to 
on-site repulsion $U$ and intrachain nearest-neighbour interaction $V$
only, we will find that additional intersite interactions
will further stabilize the BCDW.

For the 1/2-filled band, the difference between 1D and 2D is
well understood and profound. In the 1D limit, widely discussed
for polyacetylene 
\cite{reviews}, the ground state for physically relevant values
of $U>2V>0$ is a bond-order wave (BOW), with $u_j = (-1)^ju_0$.
In the isotropic 2D limit, which has been
applied to the parent compounds of the high $T_c$ superconductors,
the ground state for the same physically relevant parameters
is an antiferromagnet (AFM, {\it i.e.},
a 2k$_F$ SDW) \cite{AFM}.
Thus in the 1/2-filled band, the dominant broken symmetry depends
very strongly on dimensionality and (a) there is {\it no coexistence}
between the BOW and the SDW; and (b) the strength of the
SDW increases monotonically with the interchain hopping.

In contrast to the 1/2-filled band, broken symmetries in non-1/2-filled
commensurate correlated bands have been investigated chiefly in the
quasi-1D regime, ${t_{\perp} << t}$, and have employed primarily
Fermi surface/''k''-space arguments, which imply that finite $t_{\perp}$
{\it destroys} the Fermi surface nesting that
characterises the 1D limit, leading necessarily within the simplest
k-space interpretation to the restoration of the
metallic phase \cite{exception}.
The nesting concept, however, 
is provably applicable only in the limit of zero e-e interactions.
For strongly interacting electrons, broken
spatial symmetries are
more naturally studied by ``configuration space'' arguments
\cite{reviews,Mazumdar1}.
Using such arguments, we show below that the ground state of Eq.~(1) is a 
novel BCDW state that persists for 
{\it all} $t_{\perp}$ for the strongly interacting regime
$U \geq 4|t|$, provided that the nearest-neighbour
Coulomb interaction $V < V_c$, where $V_c$ = 2$|t|$ in the
limit $U \to \infty$ and is slightly larger for finite $U$.
For small $t_{\perp}$, the BCDW state drives a SDW
(creating a ``BCSDW'', \cite{Mazumdar3,KOY}),
but the SDW amplitude,
after reaching a maximum, vanishes as $t_{\perp}$ is further
increased. The persistence of the BCDW to the isotropic 2D
limit, its coexistence
with the SDW for small (but nonzero) $t_{\perp}$, and the vanishing of 
the SDW at large $t_{\perp}$, are all new results, quite
unexpected from established behavior of the 1/2-filled band.

We begin with the physical intuition suggesting
the existence of the 2D BCDW and the peculiar behavior of the spins.
For the 1D 1/4-filled band,
Hubbard \cite{Hubbard} showed that for large $U$
and in the presence of long-range
Coulomb interactions, there exist two distinct Wigner crystals,
represented by the configuration space $t=0$ ``cartoons''
as ...1100... and ...1010..., where the numbers denote electron
site occupancies
(for finite hopping, 1(0) actually represents site occupancy
0.5 + $\epsilon$ (0.5 -- $\epsilon$), with $\epsilon$ small but nonzero).
Ref.~\onlinecite{umt} showed that long-range Coulomb 
interactions are not essential for the Wigner
crystal ..1100.. ; this ground state is also obtained for the
1D limit of Eq.~(1), provided $V < V_c$; only for $V > V_c$ does the
the Wigner crystal ....1010.... become the ground state \cite{Hubbard}.
Importantly, for nonzero $\alpha$ or $\beta$, this ...1100...  ground state
is a {\it coexisting} BOW-CDW \cite{umt}, i.e., BCDW, as
indicated by the individual
chains in Fig.~1. The bonds
\begin{figure}[htb]
\centerline{\psfig{figure=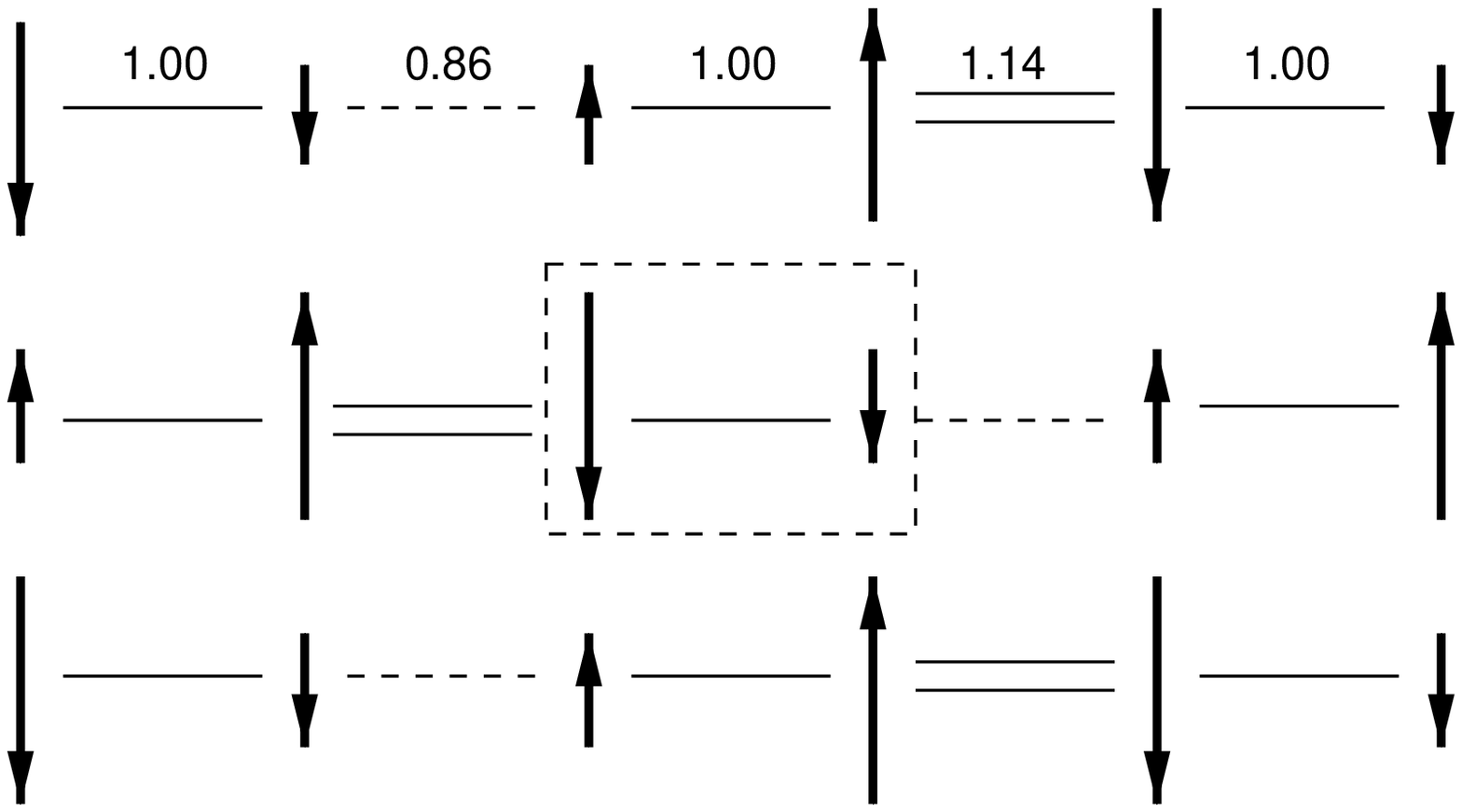,width=3.0in}}
\caption{Sketch of the novel BCDW ground state that coexists
with the SDW for small $t_{\perp}$ in the
strongly correlated, anisotropic 2D 1/4-filled band.
The arrows indicate the spin directions and their
sizes indicate the relative charge and spin densities.
The hopping integrals used to calculate
the energies of the
distorted lattices correspond to $r_4$ = 0 (see text) and
are shown above the bonds along the top chain.
This variation reflects the BOW.
Note that the charge ordering corresponds to 1D Wigner crystals along
the longitudinal, transverse, {\it and} both diagonal directions (see
text).}
\label{cartoon}
\end{figure}
in the 1D limit are singlets, but as $t_{\perp}$ is gradually increased,
the sites acquire spins \cite{Mazumdar3}, as indicated in  Fig.~1.
The spin arrangement along a chain is explained by inspection of the 1D
Wigner crystal cartoon ...1100....: each `0' is closer to a specific `1', and
therefore for finite hopping acquires the same sign (though not magnitude)
of the spin density as the site with larger charge density (since
it is the same electron that is shared between these two sites).
Our detailed numerical calculations show that a $\pi$-phase shift
between chains
gives the lowest electronic energy, implying the structure shown in
Fig.~1 for the ${t_{\perp} << |t|}$ limit.
In this Figure, the modulations of the hopping integrals are
determined by $ u_{j,m} =  u_0 [(-1)^m r_2 \cos(2k_F j - \pi/4) + r_4
\cos(4k_F j)] $, where $r_2$ and $r_4$ are the
amplitudes of the $2k_F$ and $4k_F$ distortions of $ u_{j,m}$,
respectively. Our many-body calculations show that this leads
to intrachain CDW
$\rho_c(j) \equiv \langle\sum_{\sigma}c_{j,\ell,\sigma}^{\dagger}
c_{j,\ell,\sigma}\rangle$ = 0.5+ $\rho_0 \cos (2k_F j - 3\pi/4)$, SDW
$\rho_s(j) \equiv \langle c_{j,\ell,\uparrow}^{\dagger}c_{j,\ell,\uparrow} - 
c_{j,\ell,\downarrow}^{\dagger}c_{j,\ell,\downarrow}\rangle = \rho_{s2} \cos (2k_F j - \pi/4) + \rho_{s4}(\cos(4k_F j - \pi)$, and BOW
$\langle\sum_{\sigma}B_{j,j+1,m,m,\sigma}\rangle = b_0 +  b_2 \cos (2k_F j - \pi/2) + 
b_4\cos(4k_F j - \pi)$
where the phase angles of the DW's given correspond to odd $m$, and
their amplitudes
depend on $U,V,$ and $t_{\perp}$.
 
We now discuss the consequences of increasing $t_{\perp}$ further.
Interchain hopping $t_{\perp}$ leads to partial double occupancy on a single
site
({\tiny $\uparrow$} $\downarrow$) with an energy barrier
that, while less than the bare $U$, is a $U_{eff}$ that increases with
$U$. 
For large enough $U_{eff}$, the electrons are thus confined
to their respective chains, and the BCDW state
persists
up to the isotropic 2D limit for the strongly correlated case. 
A striking feature of the BCDW state is that it is a Wigner crystal along
the chains (...1100..., periodicity 2k$_F$), transverse to the chains
(...1010..., periodicity 4k$_F$), as well as along both diagonal directions
(...1100..., periodicity 2k$_F$). The BCDW state is thus a particularly
robust broken symmetry \cite{xysym}.
For instance, by
enhancing the 4k$_F$ charge order along the transverse direction,
V$_{\perp}$ enhances the stability of the BCDW.
Similarly, the diagonal ...1100... charge ordering implies that even the
additions of hopping $t_{diag}$ and Coulomb repulsion $V_{diag}$
along the diagonals
will not destroy the Wigner crystal
for realistic parameters (in particular, $V_{diag}$ stabilizes the
BCDW relative to the Wigner crystal 
that is ...1010... along both $x$ and $y$ directions).

We next describe the evolution of the spin structure. From
the cartoon in Fig.~1, we see that for the SDW to exist it is essential that
the `0's have spin. In the small $t_{\perp}$ case, the sign of the spin 
on a `0' is 
necessarily that of the nearest intrachain `1'.
Note, however, that each
`0' also has an {\it inter}chain `1' as a neighbour and that  
for a stable
SDW the spin densities of the `1's that are neighbours of a specific `0'
must be opposite.
Therefore, with increasing $t_{\perp}$, competing effects occur. On
 the one hand, the
magnitude of the interchain exchange coupling $\sim
 t_{\perp}^2/U_{eff}$ increases. On the other hand, the spin density
 on  a site labeled `0' decreases because of
the cancelling effects of the {\it intra}- and {\it inter}-chain neighbouring
 `1's. We thus expect the
SDW to vanish at a $t_{\perp}^c$ that will depend on the magnitudes
of the bare U and V.
  
To confirm these expectations we use
exact diagonalization and constrained path
quantum Monte Carlo (CPMC) \cite{CPMC} numerical
techniques 
to calculate for representative finite 2D lattices: (i) the electronic energy
gained upon bond distortion, 
$\Delta E \equiv E(0) - E(u_{j,m})$, 
where $E(u_{j,m})$ is the electronic
energy per site with fixed distortion $u_{j,m}$ along the chains; (ii) site
charge densities $n_j$ and intrachain bond orders
$\langle\sum_{\sigma}B_{j,j+1,m,m,\sigma}\rangle$; and (iii) the z-z component
of the spin-spin correlations, for each $U$, $V$ and $t_{\perp}$.
A decreasing $\Delta E$ as
a function of $t_{\perp}$ signals the destruction of the
distortion by two-dimensionality, while a constant or increasing 
$\Delta E$ indicates a persistent
distortion \cite{reviews,Mazumdar1,umt}.
In order to determine the correct behaviour in the thermodynamic limit
from finite-size simulations, we 
choose lattices and boundary 
conditions
based on the
physical requirement that any nonzero
$t_{\perp}$ {\it must destabilise} the BCDW for {\it  noninteracting}
electrons on that 
particular finite lattice (see supplemental information).
We have studied both $r_4$ = 0 and $r_4$ = $r_2$ 
(see supplementary information), 
establishing
that while the magnitude of $\Delta E$ does depend on 
$r_4$/$r_2$, its behaviour as a function of $t_{\perp}$ does not,
provided only that  $r_2 \neq 0$. 
For brevity we present here the results for $r_4$ = 0 only. 
We note that the
magnitudes of 
$\Delta E$ for the noninteracting and the interacting cases are
very different: Coulomb interactions reduce the
$\Delta E_0 \equiv  \Delta E \mid_{t_{\perp}=0}$  
considerably. However, this merely indicates that for a given e-ph
interaction the magnitude of the distortion is less for correlated electrons
than for noninteracting electrons.
Since, however, the interacting 
single chain {\it is}
distorted \cite{umt}, 
and since our interest lies in determining the
behaviour as a 
function of $t_{\perp}$ only, the relevant quantity is not the absolute
value of $\Delta E$ but the {\it normalised} energy gained
per site upon distortion, {\it i.e.}, $\Delta E$/$\Delta E_0$. This argument
is similar to that made by Anderson \cite{Anderson} for perturbative 
treatments of coupled chains, viz., the sequence in which the intrachain 
Coulomb 
interactions and interchain hopping are included is important, and a correct
physical picture is obtained only by first including the Coulomb
interactions.

In Fig.~2(a) we compare the behaviour of $\Delta E/\Delta E_0$
for the non-interacting and interacting
($U$ = 6$|t|$, $V$ = $|t|$) cases
for three different lattices satisfying our boundary condition constraints.
Finite 4n-electron 1D periodic rings have a strong tendency to have total spin
S = 1 for nonzero U.
In the present case, the 8-site periodic ring is S = 1 in both the undistorted
and distorted states, but the 16-site ring is S = 1 in the undistorted state
and S = 0 when distorted.
The $\Delta E_0$ for the 8 $\times$ 2 and 8 $\times$ 6 then correspond to the
energy gained upon distortion by the 1D S = 1 state. This is appropriate,
since the ground states of the 8 $\times$ 2 lattice are S = 0 for the 
smallest nonzero
$t_{\perp}$, and as seen in Fig.~2(a), the exact $\Delta E$ for this case
converges smoothly
\begin{figure}[htb]
\centerline{\psfig{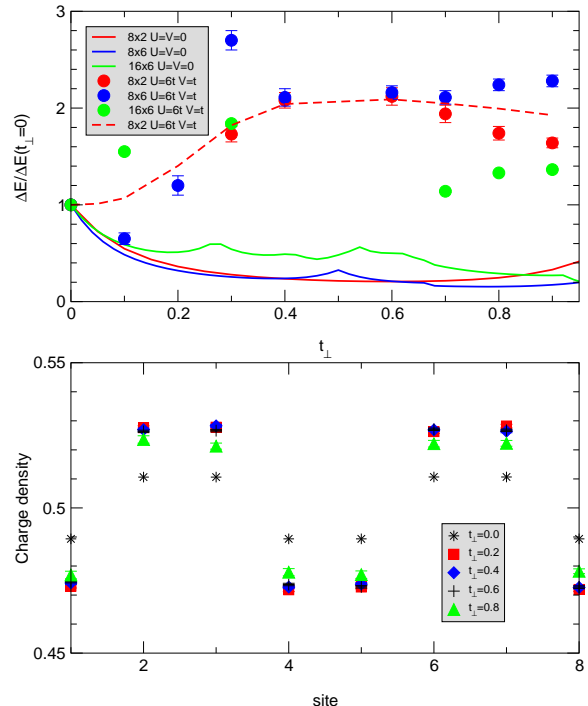}}
\caption{$\Delta E/\Delta E_0$ for the 8 $\times$ 2,
8 $\times$ 6, and
16 $\times$ 6 lattices as a function of $t_{\perp}$,
for $U$ = 6$t$, $V$ = $t$.
(b) Site charge densities on the sites
along a chain of the bond-distorted 8 $\times$ 6 lattice, for different
$t_{\perp}$ and the same $U$ and $V$ as in (a). }
\end{figure}
to the $\Delta E_0$ at $t_{\perp} \to$ 0 (the exact diagonalizations here 
included calculations for $t_{\perp}$ as small as 0.01). For the 16 $\times$ 6
lattice, the one-electron orbital occupancy at $U$ = 0 is degenerate even
for the distorted state in the region $t_{\perp}$ $>$ 0.6 indicating that
the distorted state ground state is S = 1 for large $t_{\perp}$ and nonzero 
$U$. The calculated $\Delta E$ for the 16 $\times$ 6 are therefore compared
to the triplet $\Delta E_0$ in the region $t_{\perp}$ $>$ 0.6.

As seen in Fig.~2(a),
while for the non-interacting cases the $\Delta E/\Delta E_0$
decreases rapidly with $t_{\perp}$, for the interacting cases
the $\Delta E/\Delta E_0$ either remains unchanged
or is enhanced by $t_{\perp}$.
In Fig.~2(b) we show the charge
densities on the sites along a chain for the bond-distorted
8 $\times$ 6 lattice, as a 
function of $t_{\perp}$, for the interacting case only [as expected from the
$\Delta E$ plot, the CDW amplitude decreases rapidly
for the noninteracting case]. The CDW behaviour is the
same for the 8 $\times$ 2 and the 16 $\times$ 6 lattices. Both Figs.~2(a) and
(b) indicate the enhanced stability of the BCDW in the interacting 2D case.
The same conclusion is also reached from calculations of the intrachain
bond orders. 
For the 8 $\times$ 2 lattice,
we have repeated the exact calculations also with
nonzero $V_{\perp}$, 
with the same conclusions (see supplementary information).
These results provide quantitative proof of our qualitative
arguments establishing that the BCDW is a
robust broken symmetry 
state for the interacting 2D 1/4-filled band.

In Fig.~3 we show
the interchain spin-spin correlations for the 
distorted 8 $\times$ 6 lattice for several values of $t_{\perp}$. The 
spin-spin correlations indicate a SDW 
that is enhanced between $t_{\perp}$ = 0.1
and 0.4 but then vanishes at $t_{\perp} \simeq $ 0.6.
We observe this same behaviour of the
SDW on 8 $\times$ 2 and the 16 $\times$ 6 lattices.
In all cases, the SDW amplitude initially increases, exhibits a
 maximum,
 and then vanishes
at a $t_{\perp}^c$ which decreases with the size of the system.
The initial increase of
the SDW amplitude indicates that $t_{\perp}^c$ is nonzero, a
 conclusion that is also in
\begin{figure}[htb]
\centerline{\psfig{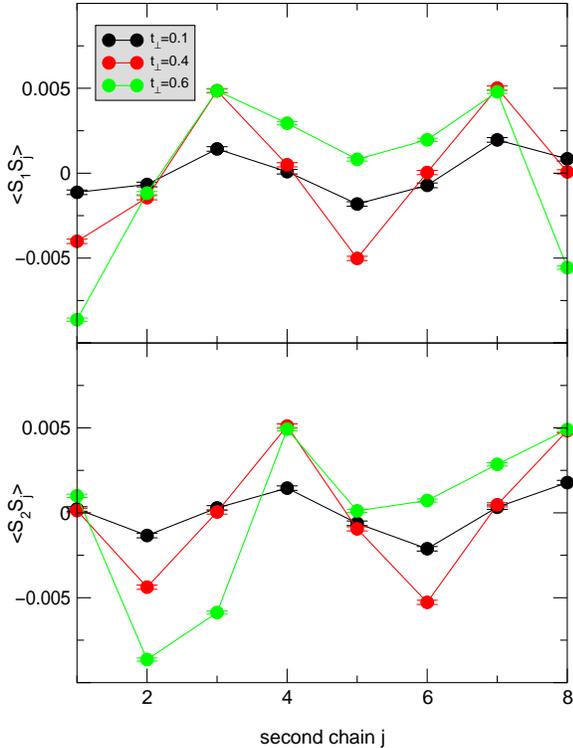}}
\caption{The z-z spin correlations between sites 1 and 2 on
the first chain of the 8 $\times$ 6 lattice and sites 1 -- 8 on the second
chain,
with $U$ = 6$|t|$, $V$ = $|t|$.
AFM correlations increase with $t_{\perp}$ up to $t_{\perp}$ = 0.4 but then
vanish at $t_{\perp} \simeq$ 0.6, even though the BCDW continues to persist
for all $t_{\perp}$ (see Fig.~2). }
\end{figure}
agreement with the experimental observation of the BCSDW state in
the weakly 2D organic
CTS \cite{Pouget,Toyota}. Based on the calculations for 16 $\times$ 6
lattice, we estimate
0.1 $<$ $t_{\perp}^c$ $<$ 0.3.
Theoretically, the 2D BCDW introduces a new mechanism for broken
symmetry in strongly
correlated lattice systems. 
The persistence of the BCDW state at large $t_{\perp}$ is related to
the Wigner 
crystal-like charge arrangement along all possible directions.
Indeed, our 1/4-filled BCDW state is remarkably similar
to the ``paired electron crystal'' state found in the continuum
electron gas by Moulopoulos and Ashcroft \cite{moulo}

Experimentally, organic CTS, which span the range
$t_{\perp} \leq 0.1t$ in (TMTTF)$_2$X to $t_{\perp}$ $\sim$ $t$ in certain
(BEDT-TTF)$_2$X, provide a critical testing
ground for our results. First, 
the 1/4-filled band
and the resulting coexistence between the BCDW and the SDW
at small $t_{\perp}$
provide a natural explanation \cite{Mazumdar3,KOY} for the
otherwise ``mysterious'' coexisting CDW/SDW state that
has been observed in (TMTSF)$_2$PF$_6$ and (TMTTF)$_2$Br \cite{Pouget}
and likely also in $\alpha$-(BEDT-TTF)$_2$MHg(SCN)$_4$ \cite{Toyota}.
Second, recent data \cite{schwartz} on frequency-dependent 
optical conductivity
in the high temperature ``metallic'' phase of
(TMTSF)$_2$ClO$_4$ and (TMTSF)$_2$PF$_6$ 
exhibit both zero- and finite-energy modes, with
the zero-energy mode having a spectral weight of only
about 1 \%.
Assuming that the (TMTSF)$_2$X are
weakly incommensurate\cite{schwartz}, our BCDW state provides a natural
explanation, with the zero-energy mode ascribed to the conductivity
of the incommensurate defects and the gap mode to the excitation
across the BCDW pseudogap that persists at high temperature
due to fluctuations associated with low dimensionality.

What might be the relation of our new BCDW state to
organic superconductivity, the mechanism for which
remains unclear despite two decades
of research?
Experimentally, in the (TMTSF)$_2$X,
the BCSDW state for small $t_{\perp}$ gives way to superconductivity
under pressure \cite{Jerome}.
Insulator-superconductor transition has also been noted in the
BEDT class of materials, where 
superconductivity is often obtained by slight modifications of the
anion (for instance,
$\alpha$-(BEDT-TTF)$_2M$Hg(SCN)$_4$ are BCSDW for $M$ = $K$, $Rb$ and $Tl$,
while the $M$ = NH$_4$
compound is superconducting; $\kappa$-(BEDT-TTF)$_2$Cu[N(CN)$_2$]X is an
antiferromagnetic semiconductor for X = Cl, but X = Br is
superconducting) \cite{Kanoda}. Both pressure and larger anions
are believed to enhance the $t_{\perp}$. One-electron
theory might suggest that the appearance of
superconductivity
is related to the
complete destruction of the background DW. 
Thus the persistence of the BCDW
state,
even in a region
where the SDW has vanished,
suggests a different scenario:
namely, interconnected or coexisting diagonal and off-diagonal long-range
order, perhaps in the presence of a few commensurability defects. Such a
scenario has been discussed by Moulopoulos and Ashcroft in their
work on the paired electron crystals \cite{moulo}, and goes back to 1970
\cite{Chester}.
Similarly, the possibility of an insulator-superconductor transition in the 
dilute 2D
electron gas, where the insulator is once again a ``paired'' Wigner crystal,
has recently been discussed \cite{phillips}. Although the 
``paired'' Wigner crystal in the 2D electron gas remains to be proved and is
presently controversial \cite{sarachik}, 
it is intriguing
that the BCDW state found here is obtained
naturally within the standard microscopic lattice Hamiltonian for organic CTS. 
Several features of our results support
the above speculation about the insulator-superconductor
transition. First,
the vanishing of the SDW and the
persistence of the BCDW,
taken together, suggest the possibility of singlet coupling between the
sites labeled `1' in Fig.~1 in this region, leading to a spin gap.  
Second, within many existing models of superconducting pairing
involving correlated
electrons, the interactions that bind two particles often also lead to
phase segregation.
In contrast, 
any binding of
commensurability defects
here would be a consequence of the
``quasi-chemical bonding''
that occurs between the ``occupied'' sites, and these pairs are
already separated. Finally, our preliminary calculations in the region 
$t_{\perp}$ $>$
$t_{\perp}^c$ in small lattices show signatures of
superconducting pairing correlations (see supplementary information), 
but these calculations will have to be
extended to larger lattices before firm conclusions can be drawn.

{\footnotesize
{\bf Acknowledgements} 
This work was supported by the NSF
and the NCSA. }

{\footnotesize Correspondence and requests for materials should be addressed
to S.M. (email: sumit@physics.arizona.edu)}


\begin{references}
\bibitem{qhe} Perspectives in quantum Hall effects: novel quantum liquids in 
low-dimensional semiconductor structures. Eds. Dassarma S. and  Pinczuk A.,
[Wiley, NY 1997].
\bibitem{hightc} Emery, V. J. and Kivelson, S. A.,
  Importance of phase fluctuations in superconductors with small
  superfluid
  density. Nature {\bf 374}, 434-437 (1995).
\bibitem{sarachik} Sarachik, M. P. and Kravchenko, S. V., Novel phenomena
in dilute electron systems in two dimensions. Proc. Natl. Acad Sci.
{\bf 96}, 5900 (1999). 
\bibitem{Anderson} Anderson, P.W., ``Confinement'' in the one-dimensional
Hubbard model: irrelevance of single-particle hopping. {\it Phys. Rev. Lett.}
{\bf 67}, 3844 (1991).
\bibitem{Pouget} Pouget, J.P., Ravy, S., X-ray evidence of charge density
modulations in the magnetic phases of (TMTSF)$_2$PF$_6$ and (TMTTF)$_2$Br.
Synth. Metals, {\bf 85}, 1523 (1997).
\bibitem{Toyota}   Sasaki, T., Toyota, N., Mysterious ground states in the
organic conductor $\alpha$-(BEDT-TTF)$_2$KHg(SCN)$_4$: mixed SDW and
CDW? Synth. Metals,
{\bf 70}, 849 (1995).
\bibitem{schwartz} Schwartz, A. {\it et al.}, On-chain
  electrodynamics of metallic (TMTSF)$_2$X salts: Observation of
  Tomonga-Luttinger liquid response. {\it Phys. Rev. B} {\bf 58}, 1261
  (1998). Vescoli, V. {\it et al.}, Dimensionality-driven
  insulator-to-metal transition in the Bechgaard Salts. {\it Science}
  {\bf 281} 1181 (1998).
\bibitem{xysym} 
The underlying $x \leftrightarrow y$ symmetry in
  the isotropic 2D limit
implies that there are two
  degenerate orthogonal 2D BCDW states for $t_{\perp}$ = $t$.
\bibitem{reviews}  
Baeriswyl, D. Campbell, D., Mazumdar, S.,
  An overview of the theory of $\pi$-conjugated polymers.
  {\it Conjugated Conducting Polymers}, H. Kiess, ed. (Springer Verlag, 1992),
pp. 7 - 134.
\bibitem{AFM} Manousakis, E., The spin-1/2 Heisenberg antiferromagnet on a square
lattice and its application to the cuprous oxides.
{\it Rev. Mod. Phys.} {\bf 63}, 1 (1991).
\bibitem{exception} A trivial exception to this rule is
  possibility of a 1/4-filled 2D 4k$_F$ CDW in which site occupancies
  alternate between 0 and 1 on adjacent sites due to very large $V$.
  This particular Wigner crystal/CDW cannot coexist with
  a BOW or SDW and is not of interest here. Furthermore, inclusion of
  long-range Coulomb interactions beyond those in Eq.~(1)
  destabilizes this Wigner
crystal relative to the one discussed in the paper.
\bibitem{Mazumdar1} Mazumdar, S. and Campbell, D.K., Broken Symmetries in the
one-dimensional half-filled band with arbitrarily long-range Coulomb 
interactions. {\it Phys. Rev. Lett.} {\bf 55}, 2067 (1985).
\bibitem{Mazumdar3} Mazumdar, S., Ramasesha, S., Clay, R.T.,
  and Campbell, D.K. Theory of coexisting charge and spin-density
  waves in  (TMTTF)$_2$Br, (TMTSF)$_2$PF$_6$ and
  $\alpha$-(BEDT-TTF)$_2$MHg(SCN)$_4$. {\it Phys. Rev. Lett.}
  {\bf 82} 1522 (1999). 
\bibitem{KOY} Kobayashi, N., and Ogata, M. Coexistence of SDW and CDW
  in quarter-filled organic conductors. {\it J. Phys. Soc. Jpn}
  {\bf 66} 3356 (1997); Kobayashi, N., Ogata, M. and Yonemitsu, K.
  Coexistence of SDW and purely electronic CDW in quarter-filled
 organic conductors. {\it J. Phys. Soc. Jpn}
  {\bf 67} 3356 (1998).
\bibitem{Hubbard} Hubbard, J. Generalized Wigner lattices in one
  dimension and some applications to tetracyanoquinodimethane (TCNQ)
  salts. {\it Phys. Rev. B} {\bf 17}, 494 (1978).
\bibitem{umt} 
  Ung, K.C., Mazumdar, S., and Toussaint,
  D. Metal-insulator and insulator-insulator transitions in the
  quarter-filled band organic conductors. {\it Phys. Rev. Lett.}
  {\bf 73}, 2603 (1994).
\bibitem{CPMC} Zhang, S, Carlson, J., Gubernatis J.E., 
Constrained path Monte Carlo method for fermion ground state. {\it Phys. Rev. B}
{\bf 55}, 7464 (1997).
\bibitem{moulo} Moulopoulos, K. and Ashcroft, N., Many-body theory
  of paired electron crystals. {\it Phys. Rev. B} {\bf 48} 11646
  (1993).
\bibitem{Jerome} Jerome, D., The physics of organic superconductors.
{\it Science} {\bf 252}, 1509 (1991).
\bibitem{Kanoda} Kanoda, K., Superconductor-insulator phase transformation
of partially deuterated $\kappa$-(BEDT-TTF)$_2$Cu[N(CN)$_2$]Br by control
of the cooling rate, Phys. Rev. B {\bf 59}, 8424 (1999).
\bibitem{Chester} Chester, G.V. Speculations on Bose-Einstein condensation
and quantum crystals, Phys. Rev. A {\bf 2}, 256 (1970).
\bibitem{phillips} Phillips, P., Sachdev, S., Kravchenko, S. and Yazdani, A.,
Quantum conductors in a plane, (cond-mat/9902025)
Proc. Natl. Acad. Sci., in press (1999).
\end{references}
\end{document}